**Mesoporous silica obtained with methyltriethoxysilane as co-precursor in alkaline medium**

Ana-Maria Putz[1], Kunzhou Wang[2], Adél Len[3], Jiri Plocek[4], Petr Bezdicka[4], Gennady P. Kopitsa[5,8], Tamara V. Khamova[5], Cătălin Ianăşi[1], Liviu Săcărescu[6], Zuzana Mitróová[7], Cecilia Savii[1], Minhao Yan[*2] and László Almásy[*2,3]

[1]*Institute of Chemistry Timisoara of Romanian Academy, Laboratory of Inorganic Chemistry, Bv. Mihai Viteazul, No.24, RO-300223 Timisoara, Romania*
[2]*State Key Laboratory Cultivation Base for Non-metal Composites and Functional Materials, Southwest University of Science and Technology, Mianyang 621010, China.*
[3]*Wigner Research Centre for Physics, Institute for Solid State Physics and Optics, Hungarian Academy of Sciences, Budapest, Hungary*
[4]*Institute of Inorganic Chemistry, Academy of Sciences of the Czech Republic, v.v.i, Husinec-Řež 1001, 25068 Řež, Czech Republic*
[5]*Institute of Silicate Chemistry of RAS, nab. Makarova 2, 199034 St. Petersburg, Russia*
[6]*Institute of Macromolecular Chemistry „Petru Poni" Iaşi, Romania*
[7]*Institute of Experimental Physics, SAS, Watsonova 47, 04001 Kosice, Slovakia*
[8]*Petersburg Nuclear Physics Institute, Gatchina, Russia*

[*]Corresponding author almasy@mail.kfki.hu and yanminhao@swust.edu.cn

**Abstract**
Mesoporous silica particles have been synthesized by sol-gel method from tetraethoxysilane (tetraethylorthosilicate, TEOS) and methyltriethoxysilane (MTES), in ethanol and water mixture, at different ratios of the of the silica precursors. Ammonia was used as catalyst at room temperature and hexadecyltrimethylammonium bromide (cetyltrimethylammonium bromide, CTAB) as the structure directing agent. Nitrogen sorption, X-ray diffraction and small-angle neutron scattering gave information on the evolution of the gel structure and pore morphologies in the function of MTES/TEOS molar ratio. Thermogravimetric and differential thermal analysis showed that with addition of MTES the exothermic peak indicating the oxidation of the low molecular weight organic fragments shift to higher temperature. A room-temperature, one-pot synthesis of MCM-41 type materials is presented, in which the variation of the MTES concentration allows to change the hydrophobicity, preserving the specific properties materials, like the ordered pore structure, large specific surface area and high porosity, making them suitable for selective uptake of guest species in drug loading applications. Specifically, the obtained materials had cylindrical pores, specific surface areas up to 1101 $m^2/g$ and total pore volumes up to 0.473 $cm^3/g$. The obtained mesoporous materials are susceptible for further functionalization to improve their selective uptake of guest species in drug delivery applications.

# Introduction

Monodisperse spherical silica particles can be synthesised by the Stöber method, starting from tetraalkoxysilane in water and alcohol as solvent, and ammonia catalyst [1]. Mesoporous silica nanoparticles with surfactant templated pores became well known since their synthesis by Kresge and co-workers in Mobil Corporation laboratories in the early 90's [2]. MCM-41 and related, surfactant templated materials are notable for their morphological properties such as large surface area, high porosity, and controllable mesopores with narrow size distribution [3]. Spherical, mesoporous MCM-41 like particles were first synthesized and reported by Grun, Lauer and Unger, modifying the Stöber method by adding the cationic surfactant [4]. Regarding the order, the regularity and the thermal stability of the structure, the MCM-41 type materials obtained in basic medium at room temperature, usually are qualitatively inferior to those obtained by hydrothermal synthesis [5]. In the ordinary conditions of temperature and pressure, lower condensation grade of the silica can be reached [6,7]. On the other hand, the low temperature sol-gel method is cheaper and advantageous as scalable technology. The presence of alcohol in the reaction mixture, as mutual solvent of water and alcoxide, facilitates homogenisation, favouring formation of spherical MCM-41 [8,9]. In surfactant templated spherical mesoporous silica, the channels are usually closed at one end, and the typical hexagonal long range order is much weaker compared to the classical MCM-41 materials [10].

For applications in adsorption and catalysis, the variation of surface hydrophobicity is required. Surface modifications of mesoporous materials are generally performed by two common ways: post-synthetic method and direct co-condensation method [11]. The latter involves a one-step co-condensation between tetraalkoxysilanes ($Si(OR)_4$) with one or more organoalkoxysilanes ($RSi(OR)_3$) through the sol-gel process in presence of a structure directing agent [12,13]. Inagaki and co-workers reported for the first time the synthesis of ordered mesoporous silica with organic groups evenly incorporated in the walls (so-called periodic mesoporous organosilicas, PMOs) using silsesquioxane precursor 1,2-bis(trimethoxysilyl) ethane as the silica source and octadecyltrimethylammonium chloride as the structure directing agent, under basic reaction conditions [14].

Organic modification of the silicate allows precise control over the surface properties and pore sizes of the mesoporous sieves and at the same time stabilizing the materials towards hydrolysis. Using organically modified precursors, like methylalkoxysilanes, the hydrophobicity of the materials can be changed in an easily controllable way. Various examples for using methyltriethoxysilane (MTES) as a hydrophobic reagent for xerogel preparation were presented in the literature attempting to achieve superhydrophobic surfaces. The character of the sol–gel derived hybrid film surface shifts from hydrophilic to superhydrophobic due to the incorporation of $CH_3$ groups in the silica surface from MTES precursor [15]. An enhancement of the surface hydrophobicity is retarding the leaching [7] and improves also the hydrothermal stability [16]. S-shape leaching curves have been obtained with an induction time increasing with the MTES content, while the more hydrophilic samples prepared without MTES exhibited immediate rapid release [16].

In a recent study, ordered mesoporous silica was synthesized at 60 ºC with NaOH catalyst, by using MTES together with TEOS, as silica precursors, in different molar ratio, without adding alcohol in the reaction mixture. It has been shown that he increased amount of MTES destroyed the mesoporous channels and the materials became amorphous at 30% MTES content [17].

In our present work, we prepared mesoporous particles, by room temperature sol-gel synthesis, using tetraethyl orthosilicate (TEOS) and methyltriethoxysilane (MTES) at various molar ratios, a cationic surfactant as templating agent, ammonia catalyst and adding alcohol into the reaction mixture. Spherical mesoporous particles of MCM-41 type have been obtained in a cost-effective way in one pot synthesis at room temperature. The hydrophobicity of the pore walls could be efficiently tuned varying the proportion of the hydrophobic silica precursor. Template removal was carried out using two methods: by calcinations and by extraction with solvent, and the morphology of the resulted materials has been characterized by several experimental techniques.

## 2. Experimental
### 2.1 Synthesis

All chemicals were commercially available: Tetraethyl orthosilicate (TEOS), (99%, for analysis, Fluka); Methyltriethoxysilane (MTES 97%, Merck); Ethanol Absolute (Riedel de Haen); $NH_4OH_{aq}$ (25%, SC Silal Trading SRL); Hexadecyltrimethyl-ammonium bromide (CTAB, Sigma).

The reactants molar ratio was: (TEOS+MTES): Ethanol: $H_2O$: $NH_3$= 1:58:144:11. The molar ratio Si:CTAB was kept constant at 0.3 and through the series where TEOS was partially replaced with MTES. Typically, 4.92 g CTAB were dissolved in a mixture of 116.64 mL distilled water and 152.4 mL of ethanol. Subsequently 37.1 mL of ammonia 25% were added; the mixtures were vigorously magnetic stirred in a closed vessel at room temperature for 30 min. Then, the sol–gel precursors (10 mL TEOS or TEOS+MTES mixed in advance) were slowly dripped (in 10 minutes) into the reaction mixture and the resulting mixture was vigorously stirred for 2 h (rotation speeds 250 rpm). The formed gel was aged for 16 h, in static condition. In the next day, the resulted suspension was several times washed with distilled water and then centrifuged at 9000 rpm (2 times, 5 minutes each and two times, 10 minutes each) until the pH of the supernatant approached neutral value. To the decanted precipitate, ca. 110 mL ethanol was added and left for 48 hours; then, the samples were dried at 40 ºC for 1 hour, and then at 60 ºC for 21 hours. These xerogel samples were labelled as Cx-60 (Table 1.).

Template removal was carried out using two methods: by calcination and by extraction with ethanol.

For calcination, a part of each dried sample has been further heated up to 550 °C and kept for 5 hours, and were labeled as Cx-550 (Table 1).

In the extraction procedure, 1g of Cx-60 samples were mixed with 200 ml of acidified ethanol (1ml of concentrated HCl: 100 ml of ethanol) under magnetic stirring (350 rot/min) for one hour, at room temperature following the slightly modified method of Du et al. [18]. Then, the samples were dried at 60 ºC for 12 hours. These samples were labeled as Cx-EtOH (Table 1).

Table 1. Sample names and synthesis conditions

| MTES / TEOS molar ratio | Drying at 60 ºC | Calcination at 550 ºC | Template extraction by ethanol |
|---|---|---|---|
| 0 / 100 | C0-60 | C0-550 | C0-EtOH |
| 10 / 90 | C1-60 | C1-550 | C1-EtOH |
| 20 / 80 | C2-60 | C2-550 | C2-EtOH |
| 30 / 70 | C3-60 | C3-550 | C3-EtOH |
| 40 / 60 | C4-60 | C4-550 | C4-EtOH |
| 50 / 50 | C5-60 | C5-550 | C5-EtOH |

**2.2. Characterization**

FTIR spectra were taken on KBr pellets with a JASCO –FT/IR-4200 apparatus.

Measurements of specific surface area of the composite silica samples were performed by low temperature nitrogen adsorption using QuantaChrome Nova 1200e analyzer. Before measurements the samples were outgassed at 120 °C for 17÷18 hours in vacuum. Determination of the surface area $S_{BET}$ (m$^2$/g) was carried out by Brunauer–Emmett–Teller (BET) method. When the BET plot produces either a too large, or negative C constant, the Langmuir surface area may be considered [19]. The total pore volume is measured at a relative pressure of $P/P_0$=0.99. The pore diameter was calculated from adsorption by BJH method. Pore size distribution was calculated from nitrogen adsorption isotherms by DFT, assuming cylindrical pore shape and NLDFT model.

Transmission Electron Microscopy (TEM) analysis have been carried out with Hitachi High-Tech HT7700 instrument, operated in high resolution mode at 100 kV accelerating voltage. Samples have been prepared by drop casting from diluted dispersions of nanoparticles in ethanol on 300 mesh holey carbon coated copper grids (Ted Pella) and vacuum dried.

Small-angle neutron scattering (SANS) measurements were performed on the *Yellow Submarine* instrument at the BNC in Budapest, Hungary [20]. Neutron wavelength of 5 and 10Å, and sample-detector distances 1.3 and 5.5 m were used; the measurements were carried out at room temperature. Preliminary SANS experiments were conducted on *Suanni* SANS instrument installed at CMRR, Mianyang, China [21]. X-ray powder diffraction (XRD) measurements were performed on Panalytical X'Pert Pro MPD diffractometer equipped with Cu-anode (Cu Kα;

λ=1.54181Å) and X'Celerator detector. Measurements were done at room temperature in transmission mode with Mylar foil in range from 0.65 to 10° in 2θ.

The thermogravimetric analysis was carried out between 25 °C and 800 °C using a 851-LF 1100-Mettler Toledo apparatus in air flow and a heating rate of 5 °C min$^{-1}$.

## 3. Results and Discussion
### 3.1. Thermal analysis

According to the TGA and DTA curves, four characteristic weight loss domains occur at: 25-150 °C; 150-280 °C; 280-380 °C; 380-800 °C. The TGA and DTA curves are presented in Supporting Information, and the relative weight losses are collected in Table 2.

The highest total weight loss (49.57%) occurs in the sample synthesised by using TEOS as the only silica precursor. Generally, with the increase of MTES content, the total weight losses are decreasing.

The first weight loss in the 25-150 °C temperature interval is accompanied by an endothermic effect on the DTA curves, due to desorption of the physisorbed and chemisorbed water [22] and of the residual solvents [23]. The weight loss in this region, decrease (from 6.8% at C0-60 sample to 4.3% at C5-60 sample) with the increase of MTES content, because the methyl groups within the framework hinder the adsorption of water.

The second weight loss in the 150-280 °C temperature interval was assigned to partial decomposition and reduction of organics to carbon [22,24], therefore to the CTAB decomposition and from the evaporation of byproducts, generated by the polycondensation of TEOS and MTES [17].

The third weight loss in the 280-380 °C temperature interval is accompanied by an exothermic effect on DTA, indicating oxidation of the low molecular weight fragments of the surfactant and methyl groups [24-27]. The positions of the exothermic maxima varied between 293.8 and 316.5 °C, shifting to higher temperature with the increase of MTES content.

For the last temperature interval (380-800 ºC), between 400 to 550 °C in case of methyl-modified silica gel, the redistribution reactions between Si–O and Si–C bonds occurred, with the loss of oligomeric siloxanes and, above 600 °C, the final densification and loss of residual atoms took place.

Table 2. Weight losses in different temperature regions

| Sample | 25-150 ºC | 150-280 ºC | 280-380 ºC | 380-800 ºC | Total weight loss |
|--------|-----------|------------|------------|------------|-------------------|
| C0-60  | 6.8%      | 28.9%      | 9.4%       | 4.4%       | 49.6%             |
| C2-60  | 5.2%      | 28.8%      | 6.5%       | 3.4%       | 43.8%             |
| C3-60  | 4.7%      | 29.1%      | 5.2%       | 3.8%       | 42.8%             |
| C5-60  | 4.3%      | 25%        | 2.3%       | 4.3%       | 36%               |

## 3.2. FT-IR measurements

All samples (xerogels, thermally treated materials and EtOH treated materials) show the characteristic vibration bands for the silica network, the hydrogen boned water molecules and the surface silanol groups [28-31]. The vibration bands of surfactant molecules are observed in the Cx-60 samples, and disappear for the ethanol-treated and calcined samples [32]. The Si-C stretching band of MTES at 1270 cm$^{-1}$ [17,33] disappears in the thermally treated samples, indicating the decomposition of methyl groups of MTES, while in the ethanol treated materials this specific peak is not changed. Overall, the IR data show the expected behavior of surfactant templated mesoporous silica materials. IR spectra and more detailed description are given in the Supporting Information.

## 3.3. Nitrogen adsorption/desorption measurements

In Figure 1 are presented the nitrogen adsorption/desorption isotherms of samples calcinated at 550 ºC and the xerogels after template extraction by ethanol. All isotherms are of IV type with no hysteresis, exhibiting sharp capillary condensation step around P/Po = 0.25 characteristic for MCM-41 materials with elongated pores [4,8,34]. For the calcined samples, the capillary condensation step is shifted toward lower relative pressure with increasing MTES content. Rabbani *et al.* reported a similar decrease of order for mesoporous silica synthesized with TEOS and MTES already at lower MTES content [17]. Thus, the ethanol added into the reaction mixtures promotes formation of ordered porosity for increased MTES content. For the ethanol-treated materials, the sharp capillary condensation step is present in all samples, proving that the ethanol added into the reaction mixture leads to pore uniformity even at high MTES concentrations. Further, the extraction with solvent is not destroying the weaker silica structure containing more methyl groups.

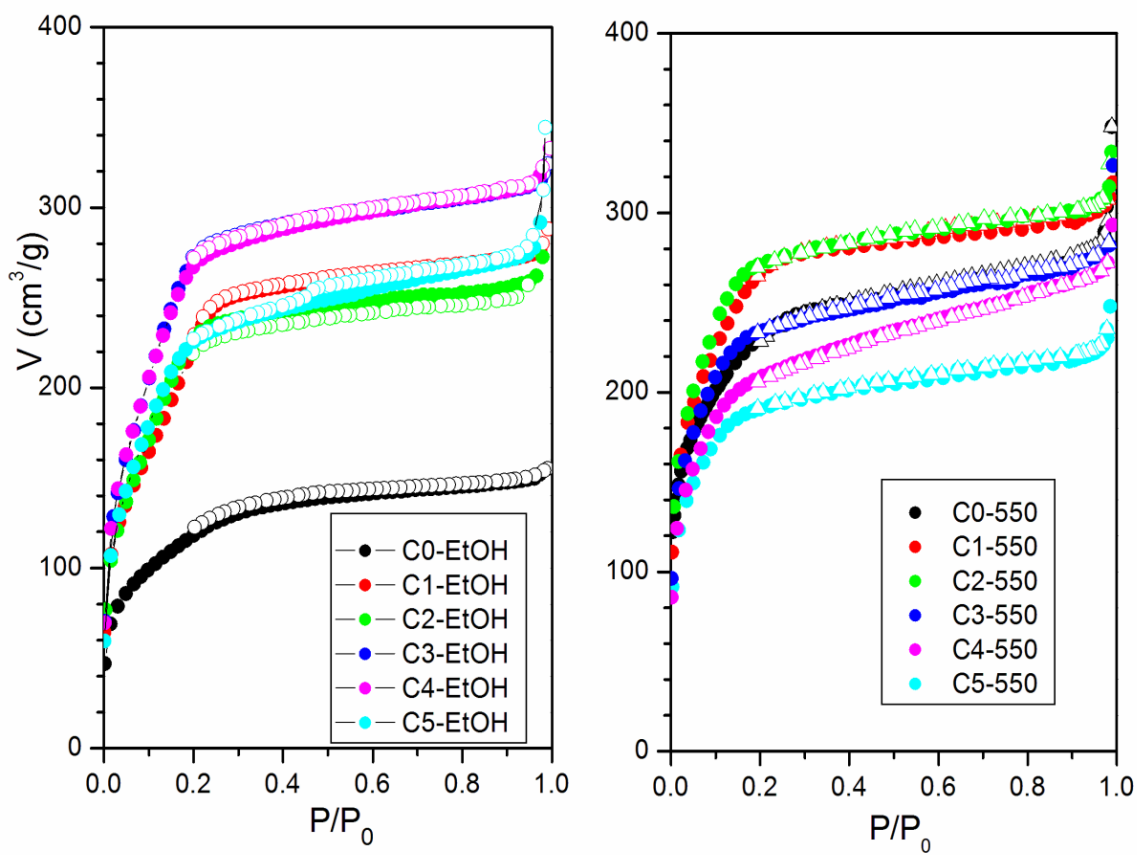

Figure 1. Nitrogen adsorption/desorption isotherms for the ethanol-treated and calcined samples.

The isotherm for C0-EtOH sample differs from the others isotherms of this series. The SSA and the total pore volume for C0-EtOH are much lower compared with the others samples, indicating that in that particular sample the surfactant extraction was not complete. Similar result was obtained in a repeated adsorption measurement. Only after further extraction steps, the surfactant could be washed out completely by ethanol. It is concluded that template extraction with ethanol is more efficient for the samples containing also MTES, while the samples made only with TEOS need longer or repeated extraction steps.

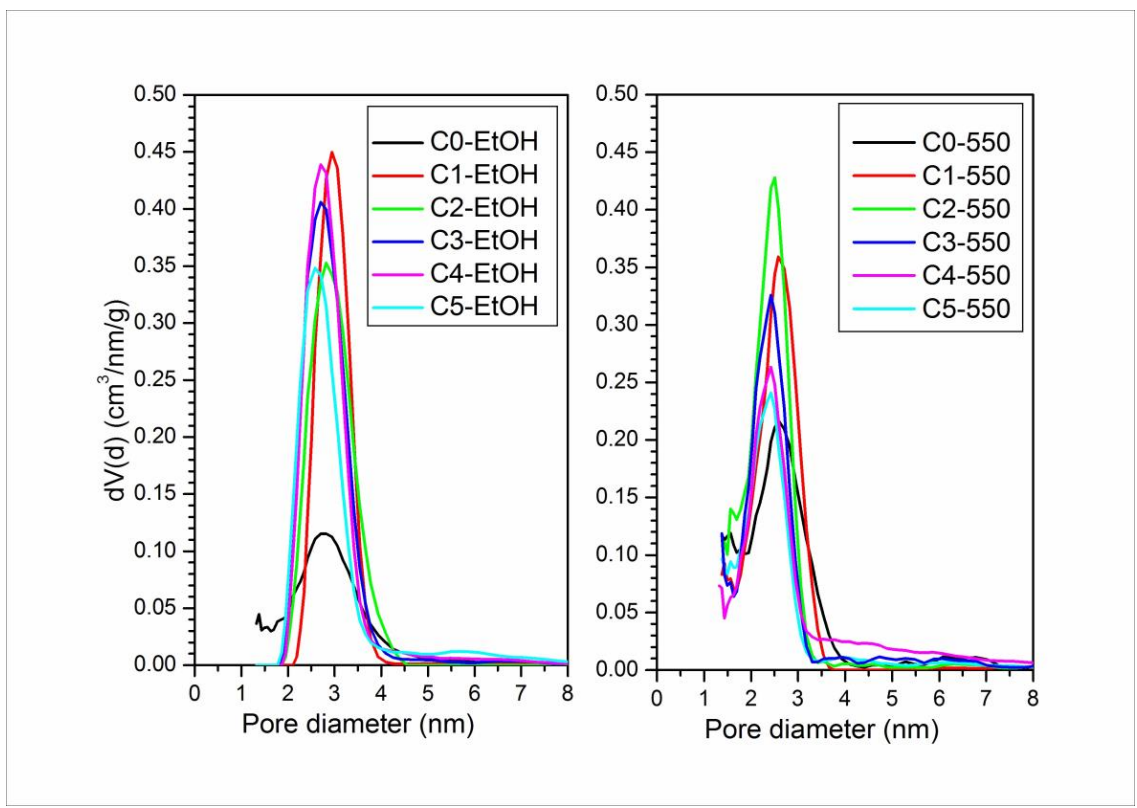

Figure 2. Pore size distributions calculated by DFT method of the samples calcinated at 550 °C

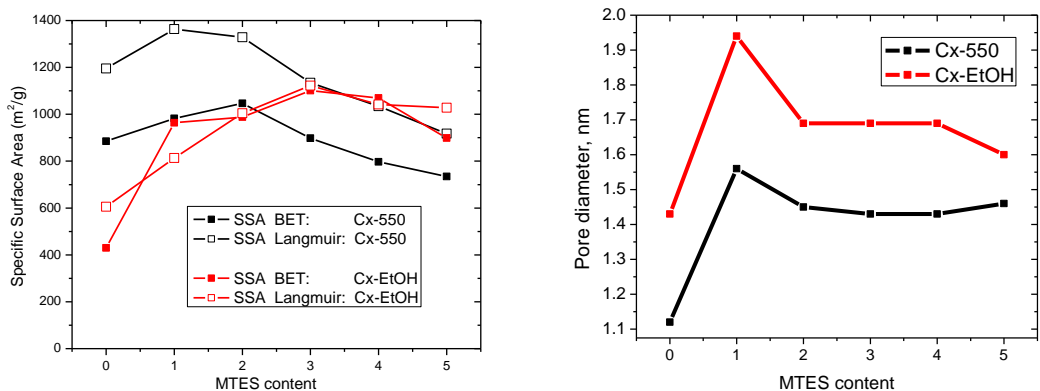

Figure 3. (a) Specific surface area and (b) BJH pore diameters (ads) in the calcinated and the EtOH treated samples.

The pore size distributions calculated by DFT are shown in Figure 2, and the development of porosity and specific surfaces with the MTES content are shown in Figure 3. The specific surface areas increase until MTES:TEOS molar ratio of 2:8 and 3:7, in Cx-550 and Cx-EtOH series, respectively. For the calcinated samples, the Langmuir analysis results in consistently higher surface areas compared to BET data, which might be related to different microporosity due the different degree of densification of the silica framework in heat treatment and solvent extraction procedures. Increasing MTES content, the pore diameters first increase, then, starting at MTES:TEOS=2:8, the pore size is nearly constant or decreasing slightly in both series, Cx-550 and Cx-EtOH (Figure 3). Decrease of pore size with increasing MTES/TEOS ratio was reported for core-shell nanoparticles with mesoporous shell, and attributed to occlusion of the pore channels as OH groups are replaced with $CH_3$ groups [16].

Nitrogen sorption data show that for small to moderate amounts of MTES, mesostructural integrity and the large pore dimensions are retained, thus allowing the variation of the hydrophobicity in a broad range. In previous studies at MTES:TEOS=2:8 molar ratio, the structural order weakened and was lost at a MTES:TEOS =3:7 [17]. In the present work, we have shown that using ethanol in the reaction mixture leads to better pore structure even at the highest tested MTES concentration.

### 3.4. Electron microscopy

Selected TEM images of the calcined samples are shown in Figure 4. The low magnification images show that all samples contain spherical particles of 200-800 nm diameters. At higher magnification, the inner porous structure of the particles is resolved, and the structure of the particle surface is visible. The comparison of several images shows that the most ordered channels are developed in the C0-550 sample (Figure 4a). At higher MTES content the porous region is visible only in the middle of the particles, while their outer part appears to be amorphous. Also, the particle surface is less sharp for the C3-550 and C5-550 samples (Figure 4c,d). This difference in the outer layer of the particles prepared with methyl containing precursors is probably due to the uneven distribution of the precursors during particle formation, since the hydrophobic, methyl containing silica atoms tend to locate near the outer surface of the particles. During calcination, the burning of the methyl groups leads to a rougher surface.

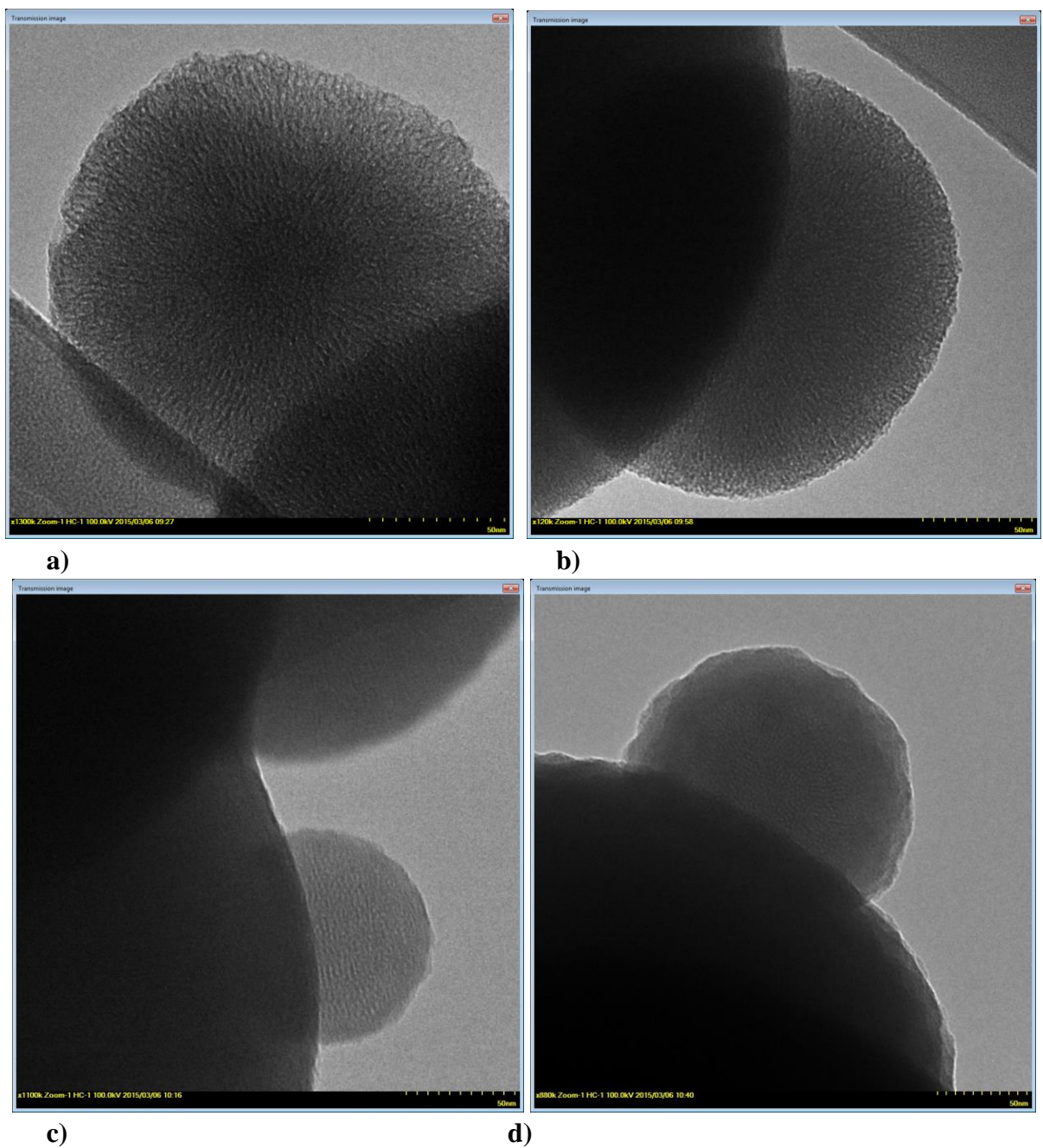

Figure 4. TEM images of silica particles: a) C0-550; b) C1-550; c) C3-550 and d) C5-550 samples. The white scale bar is 50 nm.

## 3.5. X-ray diffraction

The XRD diffractograms are shown in Figure 5. The first sharp diffraction peak is well developed in all samples, while only very weak second and third broad peaks are visible for some of them. These data indicate the locally hexagonal cylindrical pore structure, while the correlation to larger distances is weak. This is in good agreement with other studies on mesoporous silica synthesis using molar ratio TEOS: EtOH = 1:58 [35]. The peak positions in the function of the MTES content are shown in Figure 6. The gradual decrease of the lattice spacing with increasing MTES content is observed for the as-prepared, as well as for the template-removed materials. A notable feature is the asymmetric broadening of the first peak in most of the Cx-60 samples: a shoulder is apparent at 0.3 degrees to the right of the maximum. This can be explained as a concurrent cubic pore arrangement, encountered on specific mesoporous Stöber particles and induced by the sphericity of the particle that prevents the formation of long-range parallel cylinders [35,36]. After calcination at 550 °C, the peaks are broadened and this shoulder is not visible. The peak position changes from 2.4-2.6° of the xerogels to 2.8-2.9° for the calcined samples, showing shrinkage of the silica by 12% in average. In the samples treated with ethanol, the shrinkage is only Large angle XRD data contain a broad peak at about 23 degrees, which is characteristic for amorphous silica glass (see in Supporting Information).

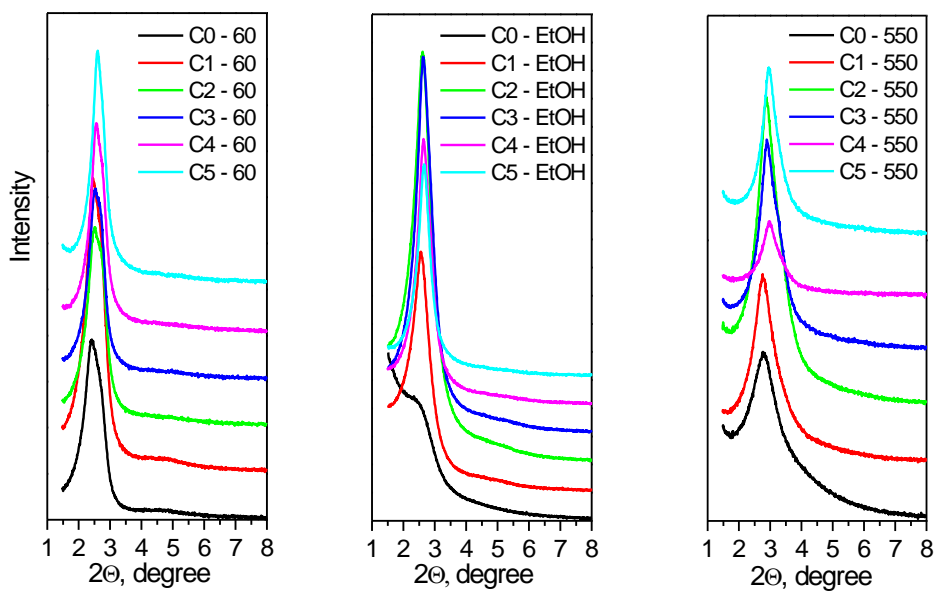

Figure 5. XRD diffraction patterns of xerogels, solvent treated and calcined samples.

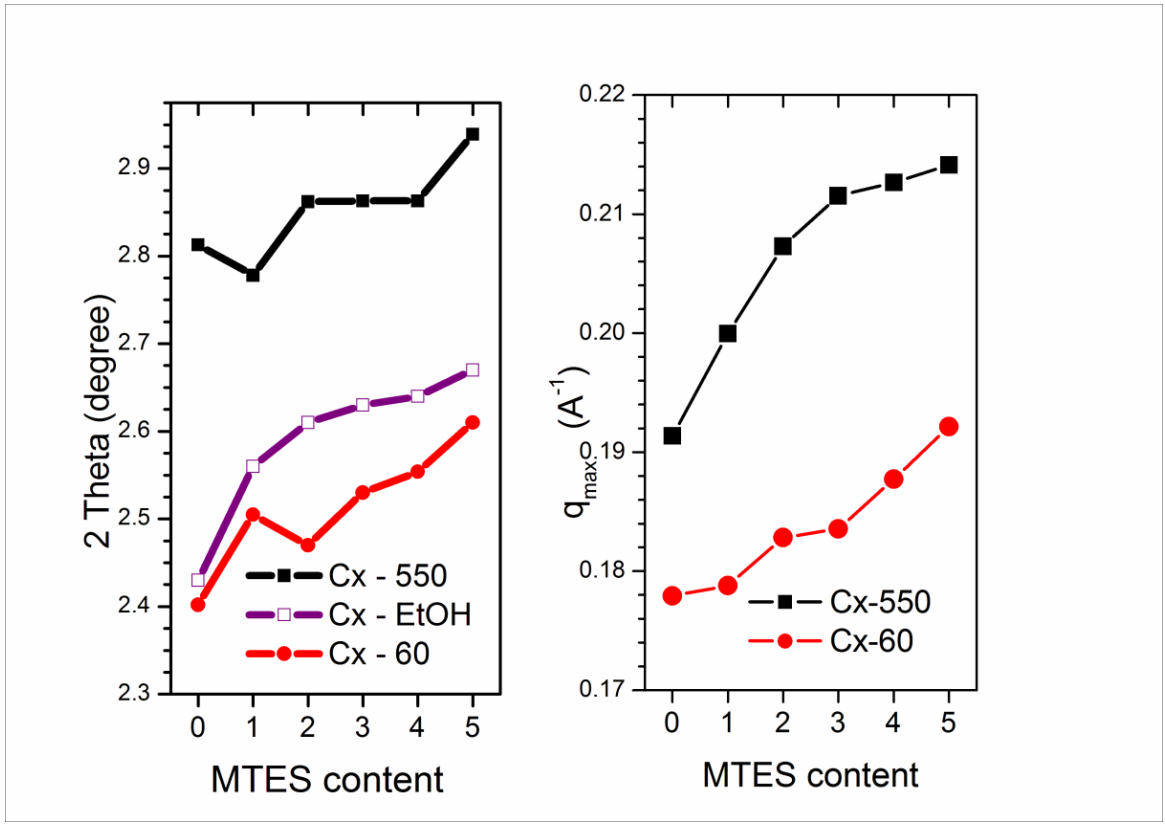

Figure 6. Position of the first XRD and SANS diffraction maximum as a function of MTES content.

### 3.6. Small angle neutron scattering

SANS scattering curves of the Cx-60 and Cx-550 sample series are shown in Figure 7. The first diffraction peak is broader compared to the XRD data due to the instrumental resolution. At small angles, the logarithmic decay of the scattering curves reflects the surface roughness of the spherical particles. An empirical model was fitted to the data, consisting of a power law decay at the small $q$ region, and a Gaussian peak. The fitting was done separately for the two $q$-ranges. The power exponent varied between -3.5 and -4.2, showing no regularity with MTES content or thermal treatment. Value of -4 would correspond to an ideally smooth interface, while the observed deviations from this value point to a rough particle surface [37,38]. The variation of the pore distance is similar to the XRD results, the trends being smoother due to the lower resolution (Figure 6).

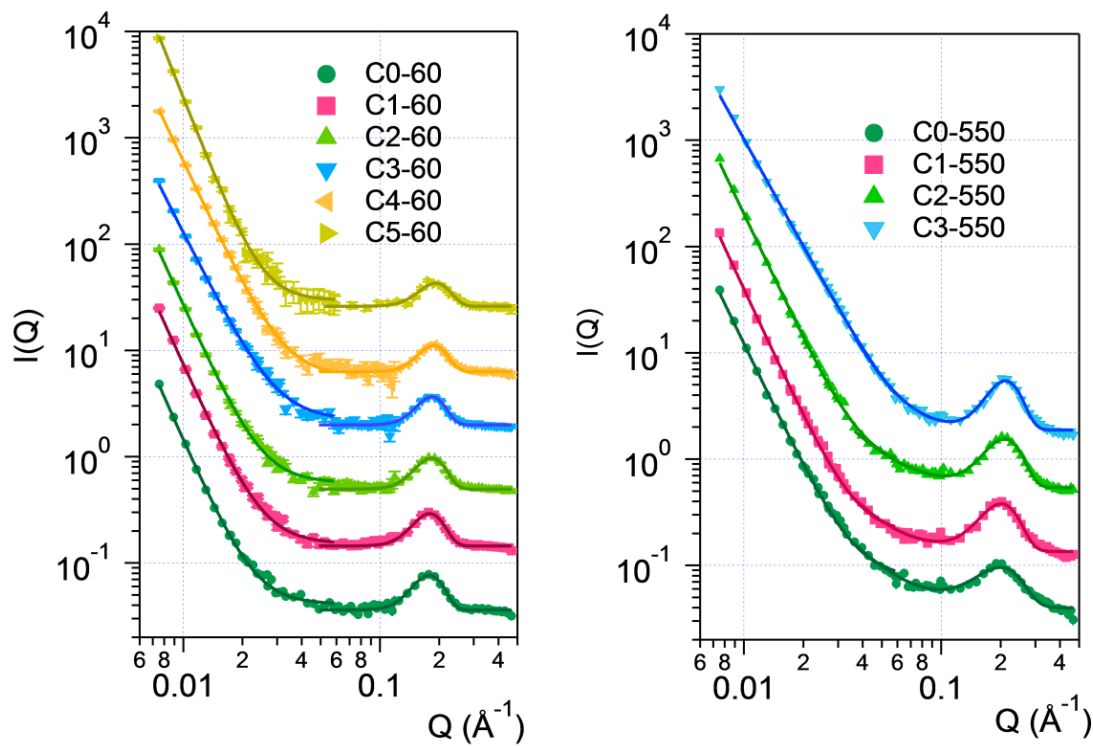

Figure 7. SANS scattering curves of the xerogels and the calcinated materials. The curves are shifted vertically for better visibility. Lines are least squares fits to power law model at low *q*, and Gaussian peak at high *q* regions.

## 4. Conclusions

A simple room temperature sol-gel method for one-pot synthesis of spherical MCM-41 type materials of varying hydrophobicity was developed, employing a mixture of silica precursors TEOS and MTES. Structural as well as thermal stability has been analysed. Electron microscopy shows spherical particles of several hundreds of nanometers, with cylindrical pores arranged mostly in radial direction. Nitrogen adsorption, XRD and SANS measurements revealed the integrity of the pore structure for all samples, and a monotonous variation of the pore size with the precursor composition. With the increase of MTES content, the numbers of $CH_3$ groups on the silica surfaces is increasing, increasing the hydrophobic character of the material. The high porosity and specific surface, as well as the monotonous changes of structural and surface parameters with varying MTES content allows tuning these parameters in a broad range for potential applications, comparatively to similar materials prepared with TEOS only.

Using ethanol as co-solvent in this room-temperature synthesis proved to be essential for formation of the mesoporous structure at high MTES concentrations, providing therefore an easy way of variation of the hydrophobicity without changing synthesis temperature.

**Acknowledgements**

Authors thank the Romanian Academy and the Inter-Academic Exchange Programs: between Romanian Academy and Hungarian Academy of Sciences and between Academy of Sciences of the Czech Republic and Romanian Academy. L. A. thanks the Hungarian and Slovak Academies for supporting a research visit to Institute of Experimental Physics, Kosice. Neutron scattering experiments from this research project has been supported by the European Commission under the 7[th] Framework Programme through the key action: Strengthening the European Research Area, Research Infrastructures. Grant Agreement N 283883-NMI3-II, and the Open Project of the Key Laboratory of Neutron Physics and Institute of Nuclear Physics and Chemistry, China Academy of Engineering Physics (No. 2014BB06).